\documentclass[10pt]{article}
\usepackage{latexsym}
\usepackage{amssymb}
\usepackage{amsmath}
\usepackage{amscd}
\usepackage{amsthm}
\usepackage[left=2cm,top=2.5cm,right=2.5cm,bottom=1.5cm]{geometry}
\usepackage[dvips]{graphicx}
\usepackage{hyperref}
\begin{document}
\begin{center}
\large{\bf{Accelerating Bianchi Type-V Cosmology with Perfect Fluid and Heat Flow in S$\acute{a}$ez-Ballester Theory}} \\
\vspace{10mm}
\normalsize{Anirudh Pradhan$^1$, Ajay Kumar Singh$^2$, D. S. Chouhan$^3$}\\
\vspace{5mm}
\normalsize{$^{1,2}$Department of Mathematics, Hindu Post-graduate College, Zamania-232 331, Ghazipur, India \\
\vspace{2mm}
$^1$e-mail: pradhan@iucaa.ernet.in; pradhan.anirudh@gmail.com }\\
\vspace{5mm} 
\normalsize{$^{3}$Department of Mathematics, School of Engineering,\\
Sir Padampat Singhania University, Bhatewar, Udaipur - 313 601, India \\
\vspace{2mm}
E-mail : ds.chouhan@spsu.ac.in}\\
\end{center}
\vspace{10mm}
\begin{abstract} 
In this paper we discuss the law of variation of scale factor $a = (t^{k}e^{t})^{\frac{1}{n}}$ which 
yields a time-dependent deceleration parameter (DP) representing a new class of models that generate a 
transition of universe from the early decelerated phase to the recent accelerating phase. Exact solutions 
of Einstein's modified field equations with perfect fluid and heat conduction are obtained within the 
framework of S$\acute{a}$ez-Ballester scalar-tensor theory of gravitation and the model is found to be in 
good agreement with recent observations. We find, for $n = 3, ~ k = 1$, the present value of DP in derived 
model as $q_{0} = -0.67$ which is very near to the observed value of DP at present epoch. We find that 
the time-dependent DP is sensible for the present day Universe and give an earmark description of evolution 
of universe. Some physical and geometric properties of the models are also discussed. 
\end{abstract}
\section{Introduction}
In the last few decades several new theories of gravitation, carefully weighed to be alternative to 
Einstein's theory of gravitation, have been developed according to an orderly plan. In alternative 
theories of gravitation, scalar tensor theories proposed by Brans and Dicke \cite{ref1}, Nordvedt 
\cite{ref2}, Wagoner \cite{ref3}, Rose \cite{ref4}, Dun \cite{ref5}, S$\acute{a}$ez and Ballester 
\cite{ref6}, Barber \cite{ref7}, Lau and Prokhovnik \cite{ref8} are most important among them. There 
are two categories of gravitational theories involving a classical scalar field $\phi$. In first category 
the scalar field $\phi$ has the dimension of the inverse of the gravitational constant $\rm G$ among which 
the Brans-Decke theory \cite{ref1} is of considerable importance and the role of the scalar field is 
confined to its effect on gravitational field equations. Brans and Decke formulated a scalar-tensor theory 
of gravitation which introduces an additional scalar field $\phi$ besides the metric tensor $g_{ij}$ and a 
dimensionless coupling constant $\omega$. This theory goes to general relativity for large values of the 
coupling constant $\omega > 500$. In the second category of theories involve a dimensionless scalar field. 
S$\acute{a}$ez and Ballester \cite{ref6} developed a scalar-tensor theory in which the metric is coupled 
with a dimensionless scalar field in a simple manner. This coupling gives a satisfactory description of 
the weak fields. In spite of the dimensionless character of the scalar field, an anti-gravity regime appears. 
This theory suggests a possible way to solve the missing-matter problem in non-flat FRW cosmologies. The 
Scalar-Tensor theories of gravitation play an important role to remove the graceful exit problem in the 
inflation era \cite{ref9}. In earlier literature, cosmological models within the framework of S$\acute{a}$ez-Ballester 
scalar-tensor theory of gravitation, have been studied by Singh and Agrawal \cite{ref10,ref11}, Ram and Tiwari \cite{ref12}, 
Singh and Ram \cite{ref13}. Mohanty and Sahu \cite{ref14,ref15} have studied Bianchi type-$\rm VI_{0}$ and Bianchi type-I 
models in Saez-Ballester theory. In recent years, Tripathi et al. \cite{ref16}, Reddy et al. \cite{ref17,ref18}, Reddy 
and Naidu \cite{ref19}, Rao et al. [20-22], Adhav et al. \cite{ref23}, Katore et al. \cite{ref24}, Sahu 
\cite{ref25}, Singh \cite{ref26}, Pradhan and Singh \cite{ref27}, Socorro and Sabido \cite{ref28} and Jamil et al. \cite{ref29} 
have obtained the solutions in S$\acute{a}$ez-Ballester scalar-tensor theory of gravitation in different context. Recently, 
Naidu et al. \cite{ref30,ref31} and Reddy et al. \cite{ref32} have studied LRS Bianchi type-II models in S$\acute{a}$ez and 
Ballester scalar tensor theory of gravitation in different context. \\

Recently, Ram et al. \cite{ref33} obtained Bianchi type-V cosmological models with perfect fluid and heat flow in S$\acute{a}$ez and 
Ballester theory by considering a variation law for Hubble's parameter with average scale factor which yields constant value of the 
deceleration parameter. In literature it is common to use a constant deceleration parameter [34-40] as it duly gives a power law 
for metric function or corresponding quantity. But it is worth mentioned here that the universe is accelerated expansion at present 
as observed in recent observations of Type Ia supernova [41-45] and CMB anisotropies [46-48] and decelerated expansion in the past. 
Also, the transition redshift from deceleration expansion to accelerated expansion is about $0.5$. Now for a Universe which was 
decelerating in past and accelerating at the present time, the DP must show signature flipping [49-51]. So, in general, the DP is 
not a constant but time variable. Recently, Pradhan et al. \cite{ref52,ref53} investigated some new exact Bianchi type-I cosmological 
models in scalar-tensor theory of gravitation with time dependent deceleration parameter.  \\

Motivated by the above discussions and observational facts, in this paper, we propose to study Bianchi type-V universe with perfect 
fluid and heat flow in S$\acute{a}$ez-Ballester scalar-tensor theory of gravitation by considering a law of variation of scale factor 
as increasing function of time which yields a time dependent DP. The out line of the paper is as follows: In Sect. $2$, the metric 
and basic equations are described. Section $3$ deals with the field equations and their quadrature solutions. The law of variation of 
scale factor is given in Sect. $4$. Subsection $4.1$ deals with the physical and geometric properties of the universe. Finally, 
conclusions are summarized in the last Sect. 5.  

\section{The Metric and Basic Equations}
We consider anisotropic Bianchi type-V line element, given by
\begin{equation}
\label{eq1} ds^{2} = dt^{2} - A^{2}(t)dx^{2} -e^{2mx}\left[B^{2}(t)dy^{2} + C^{2}(t)dz^{2}\right],
\end{equation}
where $A$, $B$ and $C$ are metric functions and $m$ is a constant. \\

We define the following parameters to be used in solving Einstein's field equations for the metric (\ref{eq1}). \\\\
The average scale factor $a$ of Bianchi type-V model
(\ref{eq1}) is defined as
\begin{equation}
\label{eq2} a = (ABC)^{\frac{1}{3}}.
\end{equation}
A volume scale factor $V$ is given by
\begin{equation}
\label{eq3} V = a^{3} = ABC.
\end{equation}
In analogy with FRW universe, we also define the generalized Hubble parameter $H$ as
\begin{equation}
\label{eq4} H = \frac{\dot{a}}{a} = \frac{1}{3}(H_{1} + H_{2} + H_{3}),
\end{equation}
where $H_{1} = \frac{\dot{A}}{A}$, $H_{2} = \frac{\dot{B}}{B}$ and $H_{3} = \frac{\dot{C}}{C}$ are directional
Hubble factors in the directions of $x$-, $y$- and $z$-axes respectively. Here, and also in what follows, a dot 
indicates ordinary differentiation with respect to t. \\

Further, the deceleration parameter $q$ is given by
\begin{equation}
\label{eq5} q = - \frac{a\ddot{a}}{\dot{a}^{2}}.
\end{equation}
We introduce the kinematical quantities such as expansion scalar ($\theta$), shear scalar ($\sigma^{2}$) and
anisotropy parameter ($A_{m}$), defined as follows:
\begin{equation}
\label{eq6} \theta = u^{i}_{;i},
\end{equation}
\begin{equation}
\label{eq7} \sigma^{2} = \frac{1}{2}\sigma_{ij}\sigma^{ij},
\end{equation}
\begin{equation}
\label{eq8} A_{m} = \frac{1}{3}\sum_{i=1}^{3}\left(\frac{H_{i}
-H}{H}\right)^{2},
\end{equation}
where $u^{i} = (0,0,0,1)$ is the matter 4-velocity vector and
\begin{equation}
\label{eq9} \sigma_{ij} = \frac{1}{2}\left(u_{i;\alpha}P_{j}^
{\alpha}+u_{j;\alpha}P_{i}^{\alpha}\right) - \frac{1}{3}\theta
P_{ij}.
\end{equation}
Here the projection tensor $P_{ij}$ has the form
\begin{equation}
\label{eq10} P_{ij} = g_{ij} - u_{i}u_{j}.
\end{equation}
These dynamical scalars, in Bianchi type-V, have the forms
\begin{equation}
\label{eq11} \theta = 3H = \frac{\dot{A}}{A} + \frac{\dot{B}}{B}
+ \frac{\dot{C}}{C},
\end{equation}
\begin{equation}
\label{eq12} 2\sigma^{2} = \left[\left(\frac{\dot{A}}{A}\right)^{2}
+ \left(\frac{\dot{B}}{B}\right)^{2} + \left(\frac{\dot{C}}{C}\right)
^{2}\right] - \frac{\theta^{2}}{3}.
\end{equation}
\section{Field Equations and their Quadrature Solutions}
The scalar tensor theories are the generalization of Einstein's theory of gravitation in which the metric is generated 
by a scalar gravitational field together with non-gravitational field (matter). We, consider the simple case of a homogeneous 
but anisotropic Bianchi type-I model with matter term with a scalar field $\phi$. Our model is based on a non-standard 
scalar-tensor theory, defined in S$\acute{a}$ez and Ballester \cite{ref6} with a dimensionless scalar field $\phi$ and 
tensor field $g_{ij}$. This alternative theory of gravitation is combined scalar and tensor fields in which the metric is 
coupled with a dimensionless scalar field. We assume the Lagrangian
\begin{equation}
\label{eq13} L = R - \omega \phi^{k} \phi_{,i}\phi^{,i},
\end{equation}
$R$ being the scalar curvature, $\phi$ a dimensionless scalar field, $\omega$ and $k$ arbitrary dimensionless constants and 
$\phi^{,i}$ the contraction $\phi_{,\alpha}g^{\alpha i}$. Here a comma ($,$) and a semicolon ($;$) stand for partial
and covariant derivative with respect to cosmic time $t$ respectively. \\

From the above Lagrangian we can establish the action
\begin{equation}
\label{eq14} I = \int_{\sum}{\left(L + 8\pi L_{m}\right)(-g)^{\frac{1}{2}}}dx^{1}dx^{2}dx^{3}dx^{4},
\end{equation}
where $L_{m}$ is the matter Lagrangian, $g$ is the determinant of the matrix $g_{ij}$, $x^{i}$ are the coordinates, $\sum$ 
is an arbitrary region of integration. When $k = 0$, our model is just the Einstein gravity with a massless minimally coupled 
scalar field coupled to gravity. By considering arbitrary independent variations of the metric and the scalar field vanishing 
at the boundary of $\sum$, the variation principle
\begin{equation}
\label{eq15} \delta{I} = 0,
\end{equation}
leads to a generalized Einstein equation 
\[
G_{ij} - \omega \phi^{k}\left(\phi_{,i} \phi_{,j} - \frac{1}{2}g_{ij}\phi_{,l}\phi^{,l}\right) = - 8\pi T_{ij},
\]
\begin{equation}
\label{eq16} 2\phi^{k}\phi_{;i}^{,i} + k \phi^{k -1}\phi_{,l}\phi^{,l} = 0,
\end{equation}
where $G_{ij} = R_{ij} - \frac{1}{2}R g_{ij}$ is the Einstein tensor; $T_{ij}$ is the stress-energy tensor of the matter
Lagrangian $L_{m}$.  \\

Since the action $I$ is a scalar, it can be easily proved that the equation of motion
\begin{equation}
\label{eq17} T^{ij}_{~ ;i} = 0,
\end{equation}
are consequences of the field equations. \\

The energy-momentum tensor is the source of gravitational field through which the effect of the perfect fluid with heat flow in
the evolution of the universe is performed. The energy-momentum tensor of a perfect fluid with heat flow has the form 
\begin{equation}
\label{eq18} T_{ij} = (\rho+p)u_{i}u_{j} - pg_{ij} + h_{i}u_{j}
 + h_{j}u_{i},
\end{equation}
where $\rho$ is the energy density, $p$ the thermodynamic pressure, $u_{i}$ the four-velocity of the fluid and $h_{i}$
is the heat flow vector satisfying
\begin{equation}
\label{eq19} g_{ij}u^{i}u^{j} = 1, 
\end{equation}
and
\begin{equation}
\label{eq20} h^{i}u_{i} = 0.
\end{equation}
We assume that the heat flow is in $x$ - direction only so that $h_{i} = (h_{1},0,0,0)$, $h_{1}$ being a function of time.
For the energy-momentum tensor (\ref{eq18}) and Bianchi type-V space-time (\ref{eq1}), the Einstein's modified field equations 
(\ref{eq16}), yield the following six independent equations as
\begin{equation}
\label{eq21} \frac{\ddot{B}}{B} + \frac{\ddot{C}}{C} + \frac{\dot{B}}{B}\frac{\dot{C}}{C} - \frac{m^{2}}{A^{2}} = 
- p + \frac{1}{2}\omega\phi^{r}\dot{\phi}^{2},
\end{equation}
\begin{equation}
\label{eq22} \frac{\ddot{A}}{A} + \frac{\ddot{C}}{C} + \frac{\dot{A}}{A}\frac{\dot{C}}{C} - \frac{m^{2}}{A^{2}} = 
- p + \frac{1}{2}\omega\phi^{r}\dot{\phi}^{2},
\end{equation}
\begin{equation}
\label{eq23} \frac{\ddot{A}}{A} + \frac{\ddot{B}}{B} + \frac{\dot{A}}{A}\frac{\dot{B}}{B} - \frac{m^{2}}{A^{2}} = 
- p + \frac{1}{2}\omega\phi^{r}\dot{\phi}^{2},
\end{equation}
\begin{equation}
\label{eq24} \frac{\dot{A}}{A}\frac{\dot{B}}{B} + \frac{\dot{A}}{A}\frac{\dot{C}}{C} + \frac{\dot{B}}{B}\frac{\dot{C}}{C} -
\frac{3m^{2}}{A^{2}} = \rho - \frac{1}{2}\omega\phi^{r}\dot{\phi}^{2},
\end{equation}
\begin{equation}
\label{eq25} m\left(2\frac{\dot{A}}{A} - \frac{\dot{B}}{B} - \frac{\dot{C}}{C}\right) = h_{1},
\end{equation}
\begin{equation}
\label{eq26} \ddot{\phi} + \dot{\phi}\left(\frac{\dot{A}}{A} + \frac{\dot{B}}{B} + \frac{\dot{C}}{C}\right) + 
\frac{r}{2\phi}\dot{\phi}^{2} = 0.
\end{equation}
The law of energy-conservation equation $T^{ij}_{~ ;j} = 0$ gives
\begin{equation}
\label{eq27} \dot{\rho} + (p+\rho)\left(\frac{\dot{A}}{A} + \frac{\dot{B}}{B} + \frac{\dot{C}}{C}\right) = 
\frac{2m}{A^{2}}h_{1}.
\end{equation}
Equations (\ref{eq21})-(\ref{eq24}) can be written in terms of $H$, $q$, $\sigma^{2}$ and $\phi$ as
\begin{equation}
\label{eq28} p = H^{2}(2q-1) - \sigma^{2} + \frac{m^{2}}{A^{2}} + \frac{1}{2}\omega\phi^{r}\dot{\phi}^{2},
\end{equation}
\begin{equation}
\label{eq29} \rho =  3H^{2} - \sigma^{2} - \frac{3m^{2}}{A^{2}} + \frac{1}{2}\omega\phi^{r}\dot{\phi}^{2}. 
\end{equation}

To solve the field equations (\ref{eq21})-(\ref{eq24}), we follow  the well established method of quadrature. For this, 
subtracting Eq. (\ref{eq21}) from (\ref{eq22}), Eq. (\ref{eq21}) from (\ref{eq23}) and Eq. (\ref{eq22}) from (\ref{eq23}), 
we get the following relations respectively:
\begin{equation}
\label{eq30} \frac{A}{B} = d_{1}\exp\left(k_{1}\int\frac{dt}{a^{3}}\right),
\end{equation}
\begin{equation}
\label{eq31} \frac{A}{C} = d_{2}\exp\left(k_{2}\int\frac{dt}{a^{3}}\right),
\end{equation}
\begin{equation}
\label{eq32} \frac{B}{C} = d_{3}\exp\left(k_{3}\int\frac{dt}{a^{3}}\right),
\end{equation}
where $d_{1},d_{2},d_{3}$ and $k_{1},k_{2},k_{3}$ are constants of integration. From Eqs.(\ref{eq30})-(\ref{eq32}), 
the metric functions can be obtained explicitly as
\begin{equation}
\label{eq33} A(t) = l_{1}a\exp\left(\frac{X_{1}}{3}\int\frac{dt}{a^{3}}\right),
\end{equation}
\begin{equation}
\label{eq34} B(t) = l_{2}a\exp\left(\frac{X_{2}}{3}\int\frac{dt}{a^{3}}\right),
\end{equation}
\begin{equation}
\label{eq35} C(t) = l_{3}a\exp\left(\frac{X_{3}}{3}\int\frac{dt}{a^{3}}\right),
\end{equation}
where
$$ l_{1} = \sqrt[3]{d_{1}d_{2}}, \indent l_{2} = \sqrt[3]{d_{1}^{-1}d_{3}}, \indent l_{3} = \sqrt[3]{(d_{2}d_{3})^{-1}},$$
$$ X_{1} = k_{1}+k_{2}, \indent X_{2}=k_{3}-k_{1}, \indent X_{3} = -(k_{2} + k_{3}),$$
and the constants $X_{1}$, $X_{2}$, $X_{3}$ and $l_{1}$, $l_{2}$,$l_{3}$ satisfy the relations
\begin{figure}[ht]
\centering
\includegraphics[width=15cm,height=10cm,angle=0]{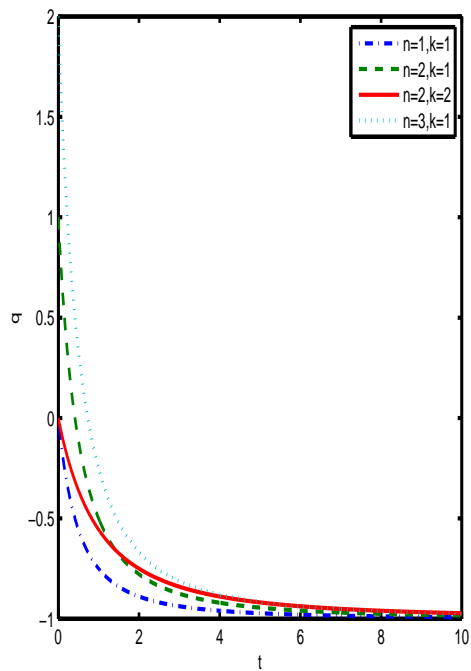} \\
\caption{The plot of deceleration parameter $q$ versus $t$.}
\end{figure}
\begin{equation}
\label{eq36} X_{1} + X_{2} + X_{3} = 0, \indent l_{1}l_{2}l_{3} = 1.
\end{equation}
The quadrature expression for the dimensionless scalar field function $\phi$, from Eq. (\ref{eq26}), is found as
\begin{equation}
\label{eq37} \phi = \left[\frac{\phi_{0}(r+2)}{2}\int\frac{dt}{a^{3}}\right]^{2/(r+2)},
\end{equation}
where $\phi_{0}$ is a constant. \\

It is clear from Eqs. (\ref{eq33})-(\ref{eq37}) that once we get the value of the average scale factor $a$, we can easily
calculate the metric functions $A$, $B$, $C$ and the scalar function $\phi$. In the next section, we are going to assume 
the value of average scale factor $a$.
\section{Law of Variation of Scale Factor $a = (t^{k}e^{t})^{\frac{1}{n}}$}
Following Yadav \cite{ref53} and Pradhan and Amirhashchi \cite{ref55}, we assume that
\begin{equation}
\label{eq38} a = (t^{k}e^{t})^{\frac{1}{n}},
\end{equation}
where $k$ and $n$ are positive constants.\\
\begin{figure}[ht]
\centering
\includegraphics[width=15cm,height=10cm,angle=0]{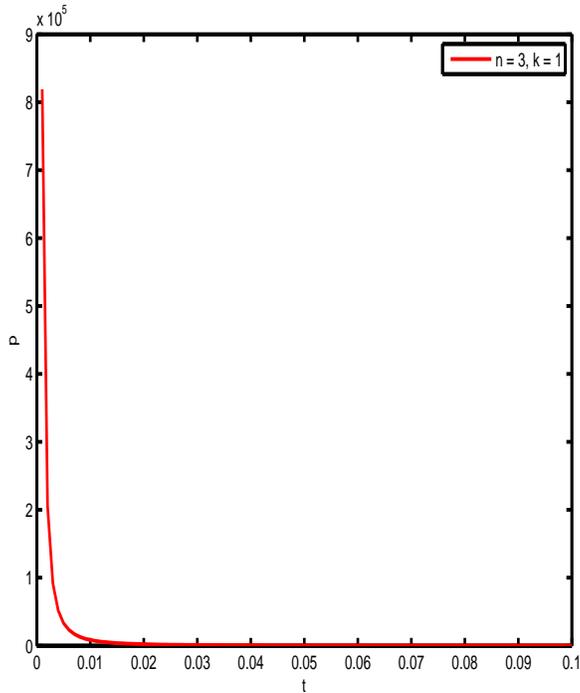} \\
\caption{The plot of isotropic pressure $p$ versus $t$.
Here $\omega = \alpha = \beta_{2} = X_{1} = m = \phi_{0} = 1$.}
\end{figure}
\begin{figure}[ht]
\centering
\includegraphics[width=15cm,height=10cm,angle=0]{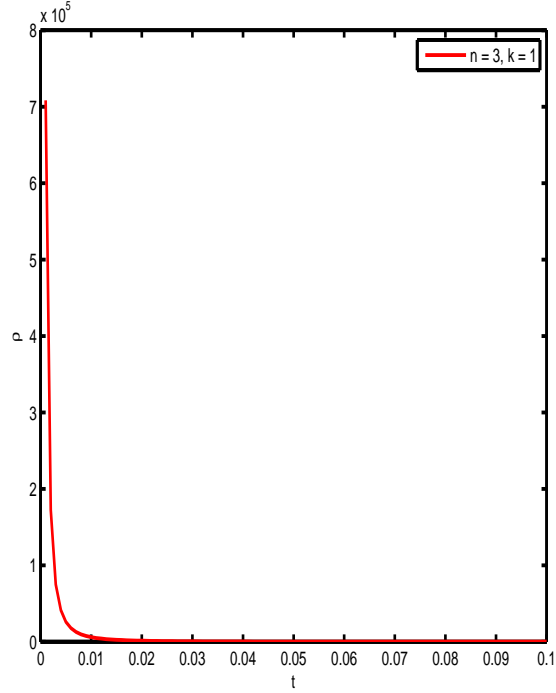} \\
\caption{The plot of energy density $\rho$ versus $t$. 
Here $\omega = \alpha = \beta_{2} = X_{1} = m = \phi_{0} = 1$.}
\end{figure}

From (\ref{eq5}) and (\ref{eq38}), we obtain the time varying deceleration parameter as
\begin{equation}
\label{eq39} q = \frac{nk}{(t+k)^{2}} - 1.
\end{equation}
From Eq. (\ref{eq39}), we observe that $q > 0$ for $t < \sqrt{nk} - k$ and $q < 0$ for $t > \sqrt{nk} - k$. It is 
observed that for $n \geq 3 ~ \& ~ k = 1$, our model is evolving from decelerating phase to accelerating phase. Also, recent 
observations of SNe Ia, expose that the present universe is accelerating and the value of DP lies to some place in the 
range $-1 < q < 0$. It follows that in our derived model, one can choose the value of DP consistent with the observation. 
Figure $1$ depicts the deceleration parameter ($q$) versus time which gives the behaviour of $q$ from decelerating to 
accelerating phase for different values of $(n, k)$ which is consistent with recent observations of Type Ia supernovae 
(Perlmutter et al. \cite{ref41}; Riess et al. \cite{ref42,ref45}; Tonry et al. \cite{ref43}; Clocchiatti et al. \cite{ref44}). \\

Using (\ref{eq38}) in Eqs. (\ref{eq33})-(\ref{eq35}), we obtain the
following expressions for scale factors:
\begin{equation}
\label{eq40} A(t) = l_{1}(t^{k}e^{t})^{1/n}\exp\left(\frac{X_{1}}{3}\int\frac{dt}{(t^{k}e^{t})^{3/n}}\right),
\end{equation}
\begin{equation}
\label{eq41} B(t) = l_{2}(t^{k}e^{t})^{1/n}\exp\left(\frac{X_{2}}{3}\int\frac{dt}{(t^{k}e^{t})^{3/n}}\right),
\end{equation}
\begin{equation}
\label{eq42} C(t) = l_{3}(t^{k}e^{t})^{1/n}\exp\left(\frac{X_{3}}{3}\int\frac{dt}{(t^{k}e^{t})^{3/n}}\right).
\end{equation}
Hence the geometry of the universe (\ref{eq1}) is reduced to
\[
ds^{2} = dt^{2} - l_{1}^{2}(t^{k}e^{t})^{2/n}\exp\left(\frac{2X_{1}}{3}\int\frac{dt}{(t^{k}e^{t})^{3/n}}\right)dx^{2} -
\]
\[
e^{2mx}\Biggl[l_{2}^{2}(t^{k}e^{t})^{2/n}\exp\left(\frac{2X_{2}}{3}\int\frac{dt}{(t^{k}e^{t})^{3/n}}\right)dy^{2} +
\]
\begin{equation}
\label{eq43}
l_{3}^{2}(t^{k}e^{t})^{2/n}\exp\left(\frac{2X_{3}}{3}\int\frac{dt}{(t^{k}e^{t})^{3/n}}\right)dz^{2}\Biggr].
\end{equation}


\begin{figure}[ht]
\centering
\includegraphics[width=15cm,height=10cm,angle=0]{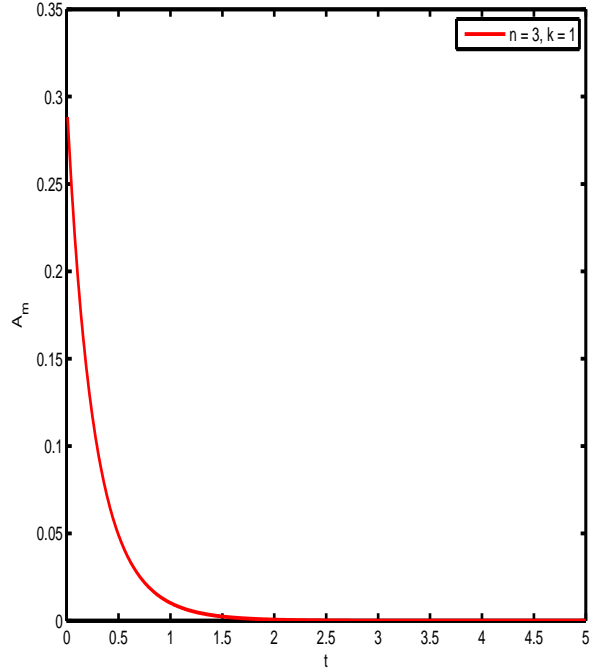} \\
\caption{The plot of anisotropic parameter $A_{m}$ versus $t$.
Here $\alpha = \beta_{2} = 1$}.
\end{figure}
\begin{figure}[ht]
\centering
\includegraphics[width=15cm,height=10cm,angle=0]{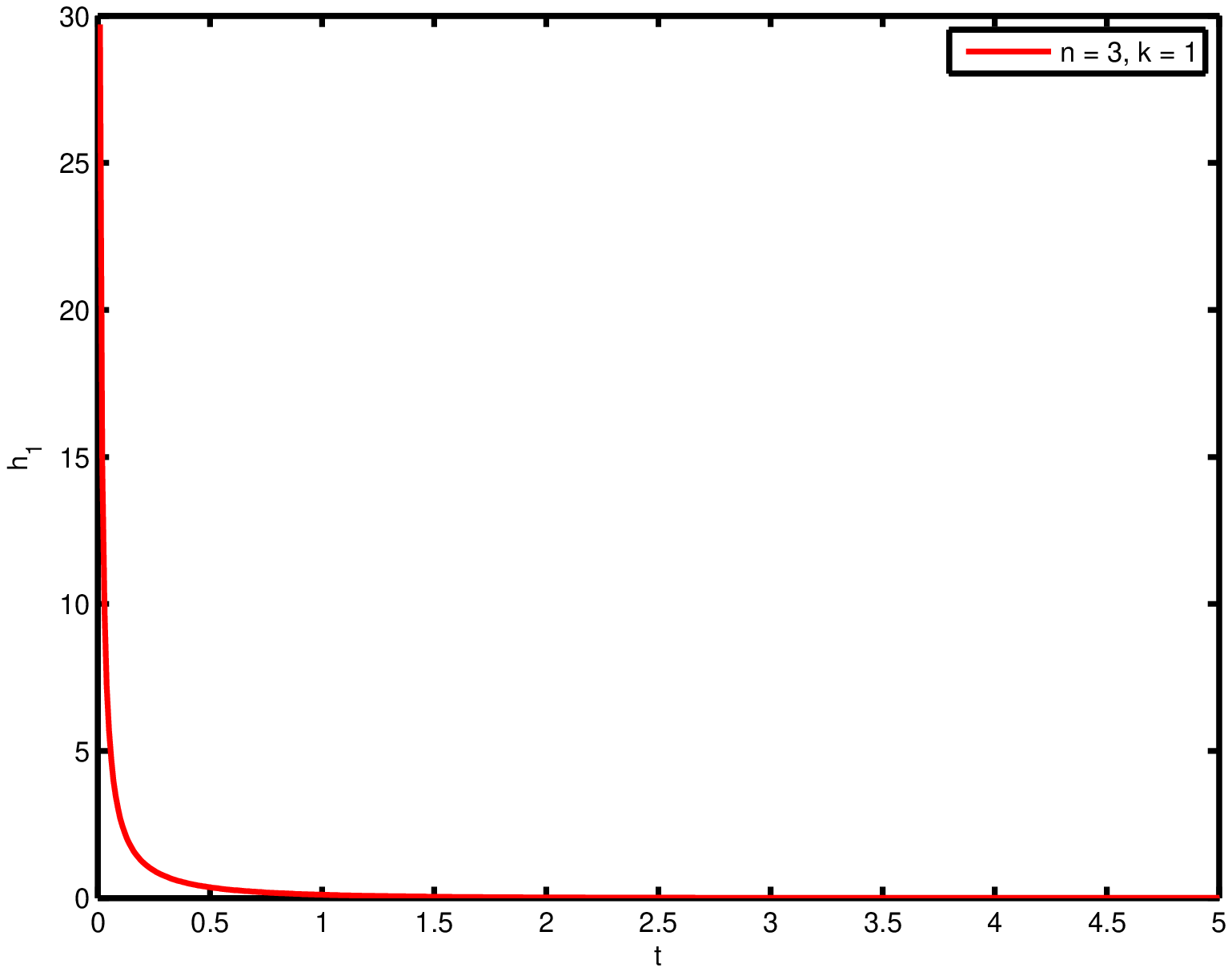} \\
\caption{The plot of heat flow function $h_{1}$ versus $t$. 
Here $\beta_{1} = m = \alpha = 1$}.
\end{figure}
\subsection{Some Physical and Geometric Properties of Model}
In derived model (\ref{eq43}), the present value of DP is estimated as
\begin{equation}
\label{eq44} q_{0} = - 1 + \frac{n}{mH_{0}^{2}t_{0}^{2}},
\end{equation}
where $H_{0}$ is the present value of Hubble's parameter and $t_{0}$ is the age of the universe at present epoch.
If we set $n = 3$ and $k = 1$ in Eq. (\ref{eq44}), we obtain $q_{0} = -0.67$. This value is very near to the  
observed value of DP (i.e., $q_{0} \approx -0.77$) at present epoch (see Cunha et al. \cite{ref56}). Hence, we restraint 
$n = 3$ and $k = 1$ in the left over discussions of the model and graphical display of physical parameters. \\\\
The solution for scalar function $\phi$, from (\ref{eq37}), is obtained as
\begin{equation}
\label{eq45} \phi = \left[\frac{\phi_{0}(r+2)}{2}\int\frac{dt}{(t^{k}e^{t})^{3/n}}\right]^{2/(r+2)}.
\end{equation}
By using the values of the metric functions from eqs. (\ref{eq40})-(\ref{eq42}) into Eq. (\ref{eq25}), 
the expression for the heat flow function $h_{1}$ is given by
\begin{equation}
\label{eq46} h_{1} = \frac{m\beta_{1}}{3(t^{k}e^{t})^{3/n}},
\end{equation}
where $\beta_{1} = 2X_{1} - X_{2} - X_{3}$. \\\\
From Eqs. (\ref{eq28}) and (\ref{eq29}) the isotropic pressure ($p$) and the energy density ($\rho$), for model (\ref{eq43}), are 
obtained as
\[
p = \frac{2k}{nt^{2}} - \frac{3}{n^{2}}\left(1+\frac{k}{t}\right)^{2} + \left[\left(\frac{1}{2}\omega\phi_{0}^{2} 
- \frac{\beta_{2}}{18}\right)\frac{1}{(t^{k}e^{t})^{6/n}}\right]
\]
\begin{equation}
\label{eq47} +\left[\frac{m^{2}}{l_{1}^{2}(t^{k}e^{t})^{2/n}}\exp\left
(\frac{-2X_{1}}{3}\int\frac{dt}{(t^{k}e^{t})^{3/n}}\right)\right].
\end{equation}
\[
\rho = \frac{3}{n^{2}}\left(1+\frac{k}{t}\right)^{2} + \left[\left(\frac{1}{2}
\omega\phi_{0}^{2}-\frac{\beta_{2}}{18}\right)\frac{1}{(t^{k}e^{t})^{6/n}}
\right] -
\]
\begin{equation}
\label{eq48} \left[\frac{3m^{2}}{l_{1}^{2}(t^{k}e^{t})^{2/n}}\exp\left
(\frac{-2X_{1}}{3}\int\frac{dt}{(t^{k}e^{t})^{3/n}}\right)\right],
\end{equation}
In view of (\ref{eq36}), it is observed that the above set of solutions satisfy the energy conservation equation 
(\ref{eq27}) identically and hence represent exact solutions of the Einstein's modified field equations 
(\ref{eq21})-(\ref{eq26}). From Eqs. (\ref{eq47}) and (\ref{eq48}), we observe that isotropic pressure $p$ and the 
energy density $\rho$ are always positive and decreasing function of time and both approach to zero as $t \to \infty$. 
Figures $2$ and $3$ depict $p$ and $\rho$, respectively, versus time $t$ showing the positive decreasing function 
of $t$ and approaching to zero at $t \to \infty$. \\   

The expressions for physical parameters such as spatial volume ($V$), directional Hubble parameters ($H_{i}, i = 1, 2, 3$), 
Hubble parameter ($H$), scalar of expansion ($\theta$), shear scalar ($\sigma$) and the anisotropy parameter ($A_{m}$) for 
model (\ref{eq43}) are, respectively, given by
\begin{equation}
\label{eq49} V = (t^{k}e^{t})^{\frac{3}{n}},
\end{equation}
\begin{equation}
\label{eq50} H_{i} = \frac{1}{n}\left(1+\frac{k}{t}\right) + \frac{X_{i}}{3
(t^{k}e^{t})^{3/n}},
\end{equation}
\begin{equation}
\label{eq51} \theta = 3H = \frac{3}{n}\left(1+\frac{k}{t}\right),
\end{equation}
\begin{equation}
\label{eq52} \sigma^{2} = \frac{\beta_{2}}{18(t^{k}e^{t})^{6/n}},
\end{equation}
\begin{equation}
\label{eq53} A_{m} = \frac{\beta_{2}n^{2}}{27}\left(1+\frac{k}{t}\right)
^{-2}(t^{k}e^{t})^{-6/n},
\end{equation}
where $\beta_{2} = X_{1}^{2} + X_{2}^{2} + X_{3}^{2}$.\\\\
From Eqs. (\ref{eq49}) and (\ref{eq51}) we observe that the spatial volume is zero at $t = 0$ and the expansion scalar 
is infinite, which show that the universe starts evolving with zero volume at $t = 0$ which is big bang scenario. From 
Eqs. (\ref{eq40})-(\ref{eq42}) we observe that the spatial scale factors are zero at the initial epoch $t = 0$ and hence 
the model has a point type singularity \cite{ref57}). We observe that proper volume increases with time. \\

The dynamics of the mean anisotropic parameter depends on the constant $\beta_{2} = X_{1}^{2} + X_{2}^{2} + X_{3}^{2} $. 
From Eq. (\ref{eq53}), we observe that at late time when $t \to \infty$, $A_{m} \to 0$. Thus, our model has transition 
from initial anisotropy to isotropy at present epoch which is in good harmony with current observations. Figure $4$ depicts 
the variation of anisotropic parameter ($A_{m}$) versus cosmic time $t$. From the figure, we observe that $A_{m}$ decreases 
with time and tends to zero as $t \to \infty$. Thus, the observed isotropy of the universe can be achieved in our model at 
present epoch. \\

It is important to note here that $\lim_{t \to 0}\left(\frac{\rho}{\theta^{2}}\right)$ spread out to be constant. Therefore the 
model of the universe goes up homogeneity and matter is dynamically negligible near the origin. This is in good agreement with 
the result already given by Collins \cite{ref58}. \\ 
   
The flow of heat along the x-direction was maximum in early universe, and it diminishes as $t \to \infty$. Figure $5$ describe the 
variation of heat flow versus cosmic time $t$ which shows the nature of $h_{1}$. From Eqs. (\ref{eq46}) and (\ref{eq52}), we also 
observe that $\frac{\sigma^{2}}{h_{1}^{2}} = $ constant which shows that shear scalar is proportional to heat conduction. 
\section{Concluding Remarks}
In this paper we have studied a spatially homogeneous and anisotropic Bianchi type-V space-time within the framework of the 
scalar-tensor theory of gravitation proposed by S$\acute{a}$ez and Ballester \cite{ref6}. The field equations have been 
solved exactly with suitable physical assumptions. The solutions satisfy the energy conservation Eq. (\ref{eq27}) identically. 
Therefore, new, exact and physically viable Bianchi type-V model has been obtained. To find the deterministic solution, we have 
considered scale factor which yields time dependent deceleration parameters. As we have already discussed in Introduction that 
for a Universe which was deceleration in past and accelerating at present time, the DP must show signature flipping [49-51] and 
so there is no scope for a constant DP. The main features of the model are as follows:  \\

$\bullet$ The model is based on exact and new solutions of Einstein's modified field equations for the anisotropic 
Bianchi type-V space time filled with perfect fluid and heat flow. \\

$\bullet$  Our special choice of scale factor yields a time dependent deceleration parameter which represents a model 
of the Universe which evolves from decelerating phase to an accelerating phase. This scenario is consistent with recent 
observations (Perlmutter et al. \cite{ref41}; Riess et al. \cite{ref42,ref45}; Tonry et al. \cite{ref43}; 
Clocchiatti et al. \cite{ref44}). \\

$\bullet$ Our whole discussions have been concentrated by restraining $n = 3, ~ k = 1$. By this choice, we find the 
present value of deceleration parameter in derived model as $q_{0} = -0.67$. This value is very near to the  
observed value of DP (i.e., $q_{0} \approx -0.77$) at present epoch (see Cunha et al. \cite{ref56}).  \\

$\bullet$ For different choice of $n$ and $k$, we can generate a class of viable cosmological models of the universe in 
Bianchi type-V space-time. For example, if we set $n = 2$ in Eq. (\ref{eq38}), we find $a = \sqrt{t^{k}e^{t}}$ which is 
used by Pradhan and Amirhashchi \cite{ref55} in studying the accelerating dark energy models in Bianchi type-V space-time 
and Pradhan et al. \cite{ref53} in studying Bianchi type-I in scalar-tensor theory of gravitation. If we set $k = 1, ~ n= 2$ 
in  Eq. (\ref{eq38}), we find $a = \sqrt{t e^{t}}$ which is utilized by Amirhashchi et al. \cite{ref59} in studying interacting 
two-fluid scenario for dark energy in FRW universe. If we set $k = 1, ~ n= 1$ in  Eq. (\ref{eq38}), we find $a = t e^{t}$ which 
is exercised by Pradhan et al. \cite{ref60} to study the dark energy model in Bianchi type-$VI_{0}$ universe. It is observed that 
such models are also in good harmony with current observations. \\

$\bullet$ It has been observed that $\lim_{t \to 0}\left(\frac{\rho}{\theta^{2}}\right)$ turn out to be constant. 
Thus the model approaches homogeneity and matter is dynamically negligible near the origin. \\

$\bullet$ We also observe that $\frac{\sigma^{2}}{h_{1}^{2}} = $ constant which shows that shear scalar is proportional 
to heat conduction (i.e. $\sigma \propto h_{1}$. \\
 
Thus, the solutions demonstrated in this paper may be useful for better understanding of the evolution of the universe 
in Bianchi type-V space-time within the framework of S$\acute{a}$ez-Ballester scalar-tensor theory of gravitation. The 
solutions presented here can be one of the potential candidates to describe the observed universe.

\section*{Acknowledgement} 
One of the authors (A. Pradhan) would like to thank the Institute of Mathematical Sciences (IMSc.), Chennai, India 
for providing facility and support under associateship scheme where part of this work was carried out. 
The authors thank Chanchal Chowla for her help in drawing the figures and also for fruitful discussions.


\end{document}